\documentclass[conference]{IEEEtran}
\IEEEoverridecommandlockouts
\usepackage{cite}
\usepackage{amsmath,amssymb,amsfonts}
\usepackage{algorithmic}
\usepackage{graphicx}
\usepackage{textcomp}
\usepackage{xcolor}
\def\BibTeX{{\rm B\kern-.05em{\sc i\kern-.025em b}\kern-.08em
    T\kern-.1667em\lower.7ex\hbox{E}\kern-.125emX}}
    
\makeatletter
\newcommand{\linebreakand}{%
  \end{@IEEEauthorhalign}
  \hfill\mbox{}\par
  \mbox{}\hfill\begin{@IEEEauthorhalign}
}
\begin{document}

\title{Using Neural Networks by Modelling Semi-Active Shock Absorber \\
}

\author{\IEEEauthorblockN{Moritz Zink}
\IEEEauthorblockA{\textit{Technische Universität Ilmenau} \\
\textit{Ilmenau, Germany}\\
moritzzink@gmail.com
}
\and
\IEEEauthorblockN{Martin Schiele}
\IEEEauthorblockA{\textit{Technische Universität Ilmenau} \\
\textit{Ilmenau, Germany}\\
martin.schiele@tu-ilmenau.de}
\and
\IEEEauthorblockN{Valentin Ivanov}
\IEEEauthorblockA{\textit{Technische Universität Ilmenau} \\
\textit{Ilmenau, Germany}\\
valentin.ivanov@tu-ilmenau.de
}
}

\maketitle

\begin{abstract}
A permanently increasing number of on-board automotive control systems requires new approaches to their digital mapping that improves functionality in terms of adaptability and robustness as well as enables their easier on-line software update. As it can be concluded from many recent studies, various methods applying neural networks (NN) can be good candidates for relevant digital twin (DT) tools in automotive control system design, for example, for controller parameterization and condition monitoring. However, the NN-based DT has strong requirements to an adequate amount of data to be used in training and design. In this regard, the paper presents an approach, which demonstrates how the regression tasks can be efficiently handled by the modeling of a semi-active shock absorber within the DT framework. The approach is based on the adaptation of time series augmentation techniques to the stationary data that increases the variance of the latter. Such a solution gives a background to elaborate further data engineering methods for the data preparation of sophisticated databases. 

\end{abstract}


\section{Motivation}
\begin{figure}[htbp]
\centerline{\includegraphics[width=8.5cm]{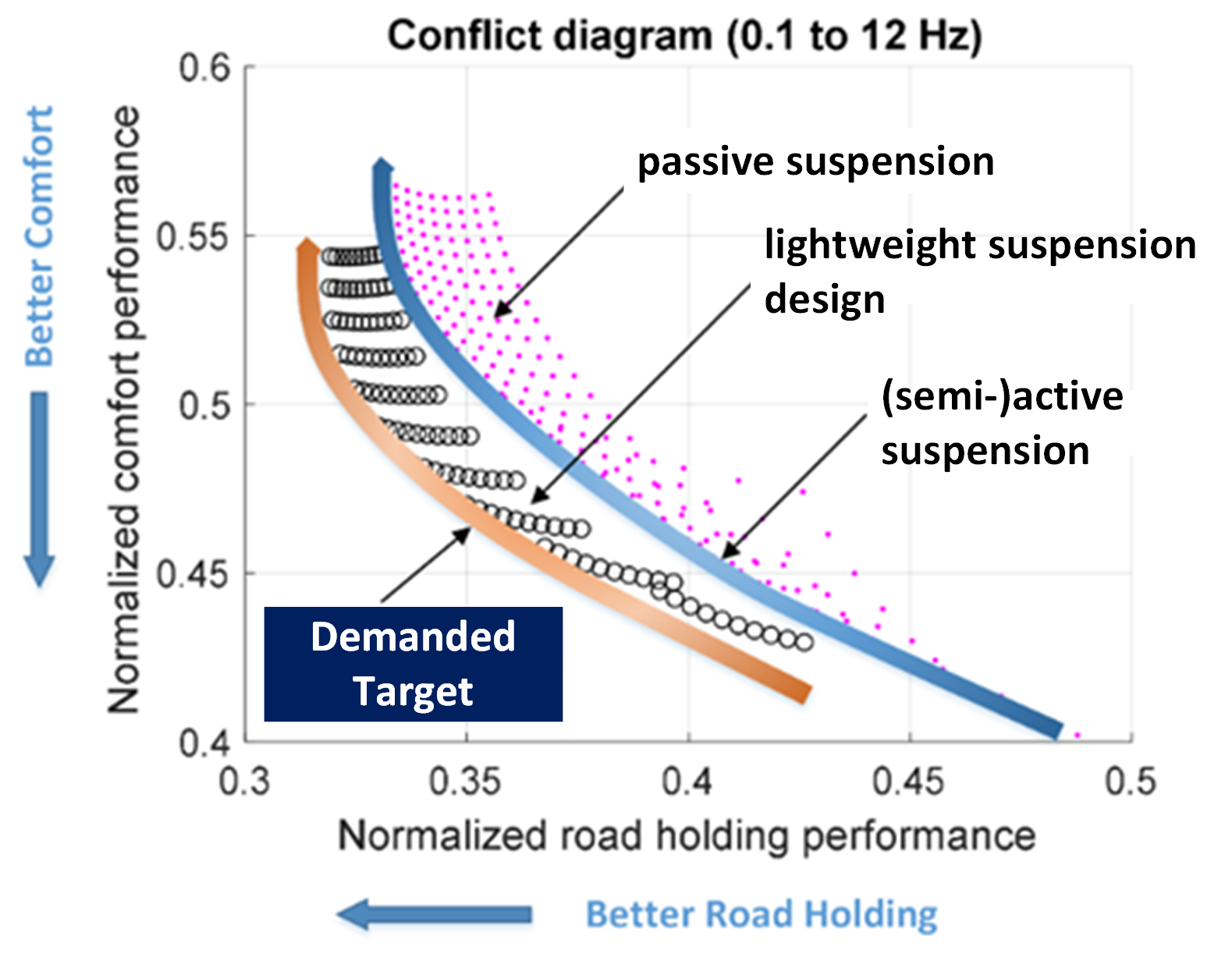}}
\caption{Trade-off of the suspension design (adapted from the project OWHEEL {https://cordis.europa.eu/project/id/872907}).}
\label{fig01}
\end{figure}

The chassis engineering of modern vehicles considers semi-active and active suspensions as an efficient tool for the improvement of the ride comfort and driving safety. Analysis of relevant research literature demonstrates that the topic of the suspension control is well elaborated from the methodological viewpoint and is characterized by variety of the available control techniques \cite{a01}, \cite{a02}, \cite{a03}, \cite{a04}, \cite{a05}. In a general case, the suspension control has a goal to achieve an optimum state, where requirements both to the road holding and ride quality will be sufficiently fulfilled. This well-known trade-off can be illustrated with Fig.~\ref{fig01}.  

The (semi-)active suspension design implies sophisticated strategies requiring a thorough parameterization of the controller with considerable computational and time efforts. In this regard, a certain support can be provided by using artificial intelligence (AI) tools that can be confirmed with results of some recent studies, e.g. \cite{a06}, \cite{a07}, \cite{a08}, \cite{a09}. The presented work will demonstrate the use of AI technique based on the reinforcement learning (RL). 

\begin{figure}[htbp]
\centerline{\includegraphics[width=8.5cm]{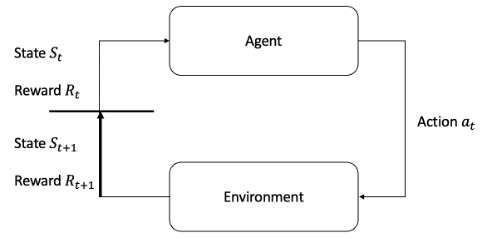}}
\caption{Principle of Reinforcement Learning.}
\label{fig02}
\end{figure}

The RL uses concept of an agent, which can operate autonomously in an environment and can learn from its feedback \cite{a10}, Fig.~\ref{fig02}. Therefore, an adequate environment must be provided for the agent. It should be noted that the real-time training of the agent on the real automotive system can be difficult due to the following potential reasons:
\begin{itemize}
\item risk of accident;
\item very long test duration;
\item limited data availability for the agent without previous performing of huge number of driving maneuvers.
\end{itemize}

To ensure an efficient way for providing training samples to the agent, the training environment can be built based on a digital twin with NN \cite{a11}. Although it is possible to use a NN for the controller design directly \cite{a12}, an advanced approach can include elements of an autonomous self-learning control system \cite{a13}. In this way, the system behavior of the shock absorber can de facto be emulated offline. In addition, digital twins can represent derived quantities that can only be defined with additional measuring systems, which are not available on mass-production cars. This procedure has been used in the presented study to directly detect critical wheel load fluctuations and to prevent them at an early stage through targeted adaptation of the shock absorber characteristic curve.

\section{Modeling shock absorber using a neural network}
Various principles are available for the technical realization of semi-active and active shock absorbers, such as electromechanical, electro-hydraulic, magnetorheological and electrorheological systems. The absorber analyzed in this study corresponds to the design outlined in \cite{a03}. In the considered case, the absorber realizes the damping force \(F_{D}\) as function of the absorber velocity \(\dot{\Delta}_{Displ}\) via the pressure modulation of a proportional valve, which changes the flow resistance and, therefore, the damping parameter \(k_{D}\):

\begin{equation}
F_{D}={k_D}\cdot f(\dot{\Delta}_{Displ}).\label{eq01}
\end{equation}

In the present system, the principle of the adjustable proportional valve was used, which can be continuously varied by the applying control current.

For the simulation purposes, the generic environment from Fig.~\ref{fig02} has to be replaced by a data-driven model of the shock absorber, a so-called digital twin, which mirrors exactly the behavior of the technical system. At that a neural network is used as modeling tool. If the shock absorber operation is represented as an unknown function \(f(x)=y\), where \(x\) is for the input variables of the system "shock absorber" as a vector and \(y\) is for the change of the rod position \({\Delta}_{Displacement}\), then the model must approximate this function:

\begin{equation}
f(x)={{\Delta}_{Displacement}}.\label{eq02}
\end{equation}

The input vector \(x\) is provided to the model to form examples in a sequence \(S\) together with the correlating output variable, which are then fed into the network \cite{a09}:

\begin{equation}
S=[(x_1,y_1),...,(x_m,y_m)]\in(X \times Y)^m.\label{eq03}
\end{equation}

In the considered case, the input variable vector is described by the following entries:
\begin{itemize}
\item Vehicle velocity, \(V\), [km/h];
\item Current, \(I\), [A];
\item Rod position, \(Displacement\), [mm];.
\end{itemize}

\begin{figure}[htbp]
\centerline{\includegraphics[width=8.5cm]{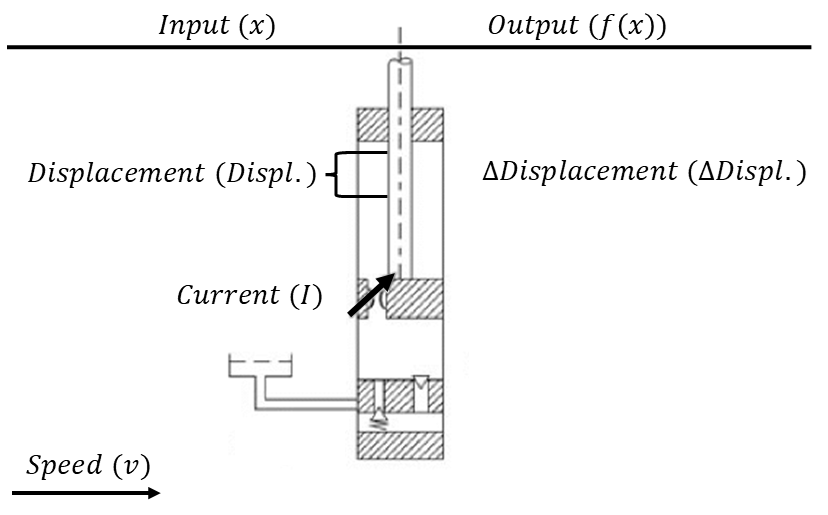}}
\caption{Input Parameters for Prediction Procedure.}
\label{fig03}
\end{figure}

The goal is now to represent the unknown function by the input variables just mentioned by the neural network, which are also explained on Fig.~\ref{fig03}:

\begin{equation}
f(V,I, Displacement) \approx {{\Delta}_{Displacement}} = \hat{y}.\label{eq04}
\end{equation}

\begin{figure}[htbp]
\centerline{\includegraphics[width=8.5cm]{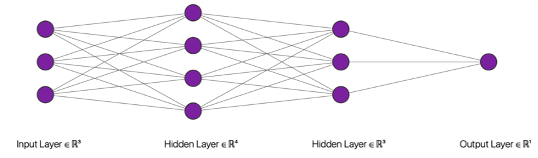}}
\caption{Neural Network Architecture for the Regression Task.}
\label{fig04}
\end{figure}

Although there are many complex architectures and ways to build neural networks, the presented case is relatively simple but sufficient for the purposes of the study. To build a regression model, fully connected neurons can be used, which are then cascaded into layers and build a multi-layer perceptron (MLP). For a continuous value to be predicted, only one neuron is used in the output layer, making \({\hat{y}}\in\mathbb{R}\), Fig.~\ref{fig04}.  

\section{Data}
 Because only one semi-active shock absorber is being discussed, a quarter vehicle model is basically sufficient for the purposes of this study. At that the vehicle pitch and roll dynamics is not considered. The data provision for training the corresponding neural network has been done on the basis of real driving tests on the experimental vehicle with parameters from Table~\ref{tab1}. The vehicle has been instrumented with the necessary sensors to measure all relevant variables as the current and the displacement. The vehicle velocity estimator has also been included for mapping the previously defined variables from (\ref{eq04}).  
 
\begin{table}[htbp]
\caption{Vehicle Specification}
\begin{center}
\begin{tabular}{|c|c|}
\hline
Vehicle type & all-wheel-drive sport utility vehicle (SUV)\\
\hline
Curb weight & 2275 kg \\
\hline
Tires & 235/55 R19 \\
\hline
Wheel base & 2660 mm \\
\hline
Track width & 1625 mm \\
\hline
\end{tabular}
\label{tab1}
\end{center}
\end{table}
 
\begin{figure}[htbp]
\centerline{\includegraphics[width=6cm]{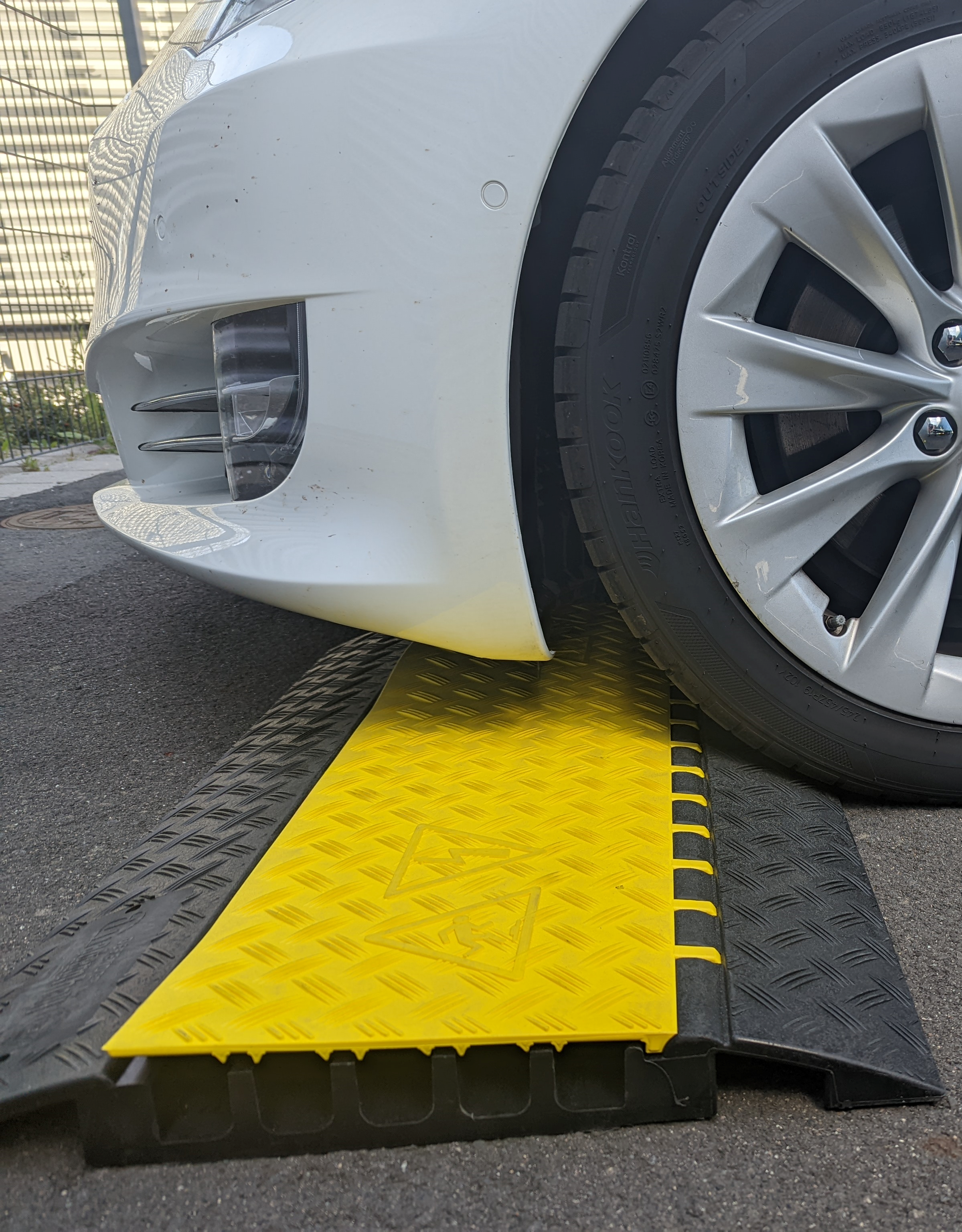}}
\caption{Driving Test with Obstacle.}
\label{fig05}
\end{figure}

During the road tests the vehicle was driven over a fixed obstacle, Fig.~\ref{fig05}, with the variation of the shock absorber current and the vehicle velocity. In addition, a test was carried out in a stationary position, in which the force was transferred to the vehicle body at a constant current. The table outlining the test program is given in Appendix. After recording the data, a scaling (\ref{eq05}) is performed to transfer all the variables to a range of values of \(\mathbb{W}=[0,1]\), following the approach from \cite{a12}. This step is required to ensure a more stable training of the model \cite{a14}.

\begin{equation}
x_{scaled}=\frac{x-x_{min}}{x_{max}-x_{min}}.\label{eq05}
\end{equation}

To get a better understanding of the inherent data values, the original data can be found feature-wise in the Appendix. Fig.~\ref{fig06} shows a clearly noisy image of the previously scaled data.  Furthermore, Fig.~\ref{fig07} shows for one particular experiment what happens in the moment when the obstacle is hit at the specific velocity with the selected current (Fig.~\ref{fig07} is the red marking on Fig.~\ref{fig06}). The approach to apply the processing steps is discussed next.

\begin{figure}[htbp]
\centerline{\includegraphics[width=8.5cm]{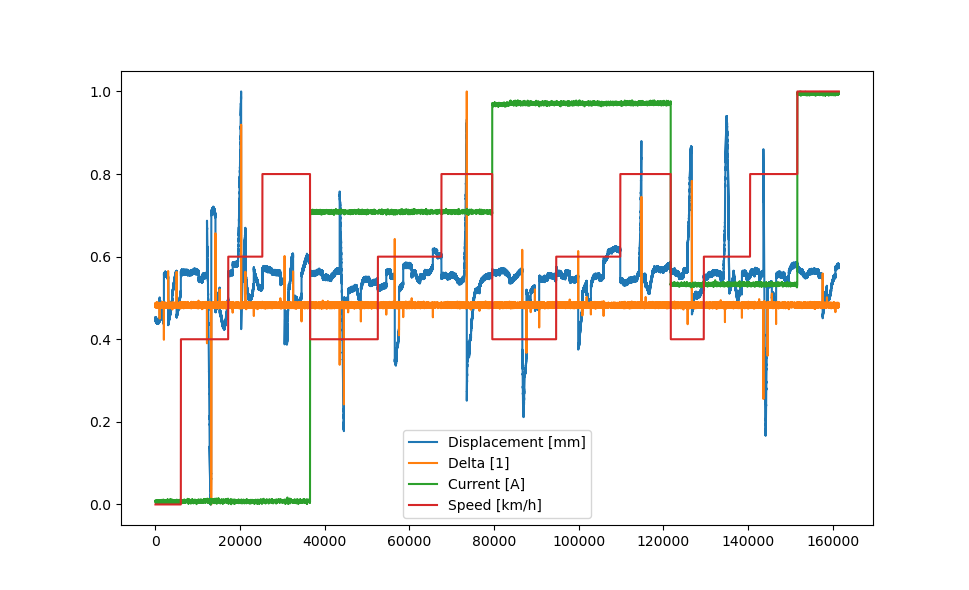}}
\caption{Scaled Noisy Data.}
\label{fig06}
\end{figure}

\begin{figure}[htbp]
\centerline{\includegraphics[width=8.5cm]{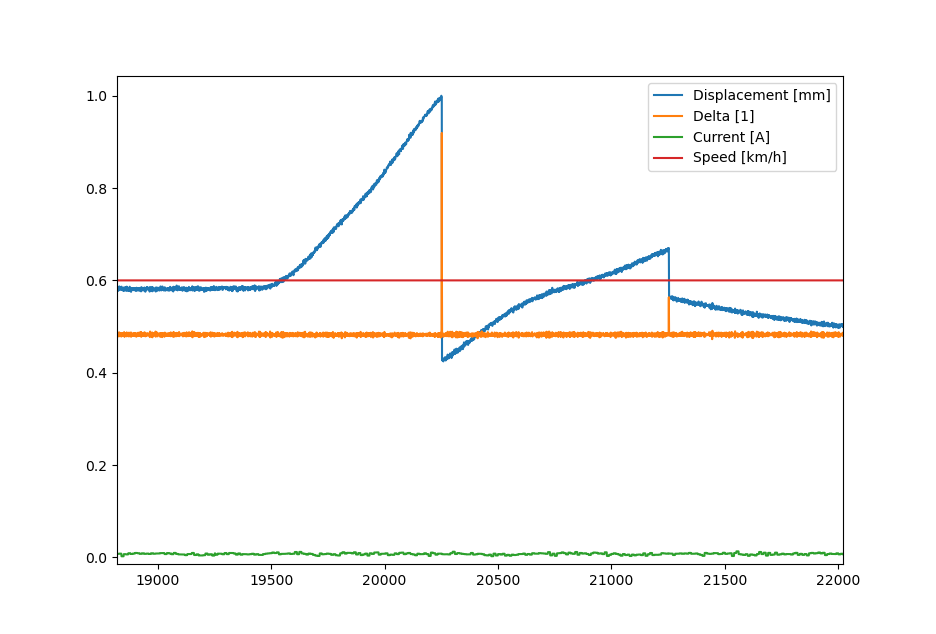}}
\caption{Shock Absorber Reaction after Obstacle Was Hit.}
\label{fig07}
\end{figure}

\section{Indexing and augmentation results}

It can be identified from Fig.~\ref{fig06} and Fig.~\ref{fig07} that about 165,000 data points were collected. However, the most of them are without any information content, as they simply fluctuate around a constant value. This is due to the very high sampling rate of the sensor system of 1000 Hz, which could not be changed in the stored vehicle model. With these data, it is impossible to obtain meaningful results of the model: because of the dominance of the noise, only its mean value would be learned. As the absolute values must not be changed, the application of a moving average is not a sensible alternative since a compression of the raw data influences it to a not inconsiderable extent. 

\begin{figure}[htbp]
\centerline{\includegraphics[width=8.5cm]{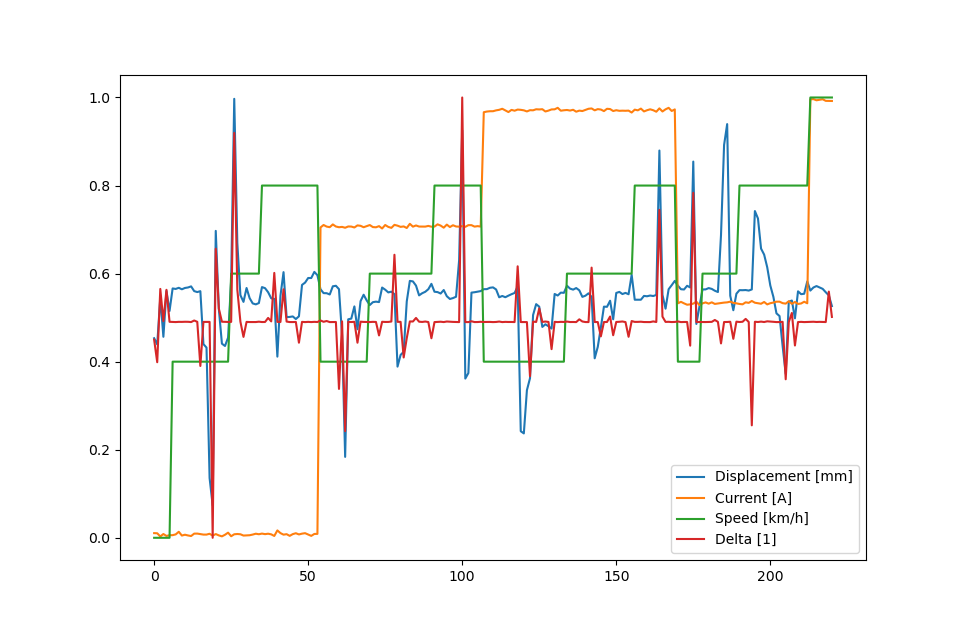}}
\caption{Cleaned Data by Indexing.}
\label{fig08}
\end{figure}

To address the mentioned problems, an approach is proposed to extract relevant, meaningful features. For this purpose, the \("Indexing"\) procedure has been developed. A proprietary filtering process is being developed for this purpose. Only those data were selected, which also have the necessary information content, i.e., for which the \({\Delta}_{Displ}\) value lies above/below a certain limit. Afterwards the index of these output \(y\)-data is being stored. The input variables (\(x\)-data) with the same index are selected and all others were deleted. This way a data set can be generated without containing any empty information. This ensures, if the generalization is sufficiently good, that the model can be interpolated between the highest and lowest data points. However, after applying \("Indexing"\) procedure, there are too few data points left to train that would lead to extreme overfitting during the training, Fig.~\ref{fig08}.

To initially increase the amount of data and to maintain the relative significance, the technique of so-called \("Data~Augmentation"\) has been used. This usually involves making geometric or non-geometric changes to images or extending time series data using various techniques \cite{a15}. At this point it was decided, inspired by images and time series, to add a Gaussian noise to the data to increase the variance of these:

\begin{equation}
p\left ( x \right )=\frac{1}{\sqrt{2\pi \sigma ^{2}}}e^-{\frac{\left ( x-\mu  \right )^{2}}{2\sigma ^{2}}}.\label{eq06}
\end{equation}

\begin{figure}[htbp]
\centerline{\includegraphics[width=8.5cm]{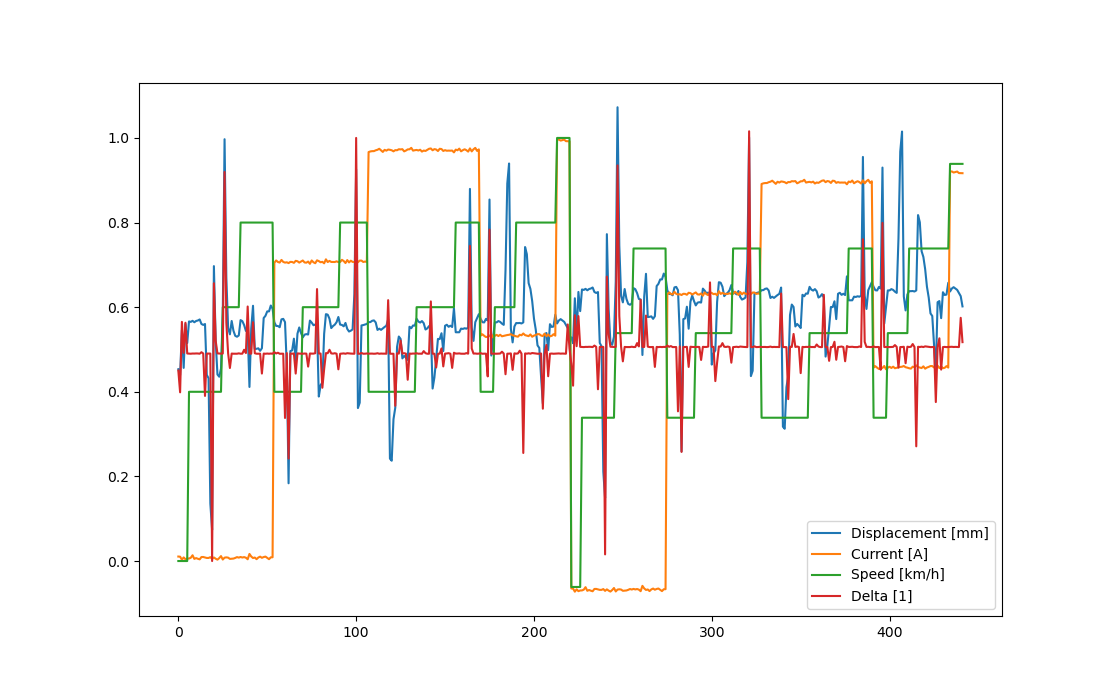}}
\caption{Cleaned and Augmented Data.}
\label{fig09}
\end{figure}

In \ref{eq06}  the parameter $\mu$  represents the mean value and $\sigma$ the standard deviation, with the values for $\mu$=0 and 
$\sigma$=0.05 chosen in the experiments. This noise was used and applied to the original data set. Subsequently, two resulting data sets (original and augmented) were merged. The data available after this procedure are shown in Fig.~\ref{fig09}.

\section{Oversampling and regression tasks}

An unbalanced data set often leads to poor results both in classification and regression tasks, because the most frequently occurring classes are highly overrepresented and thus, first, \emph{are fed to the model more often during training and, second, attempts are made to reduce the error with respect to the most frequently occurring class} \cite{a16}.

However, the probability distribution of the values of the training data do not need to correspond to those in the "real" world. For regression tasks in the technical environment, where each state is to be approached about the same number of times, the aim is to have a data set as balanced as possible.

While there are some established approaches for unbalanced classification problems (SMOTE, ClusterCentroids etc.), the selection of the procedures for solving regression problems is a non trivial task. At this point it was decided to compute a histogram of the available data and to perform an "oversampling" based on the class of values, which are most frequently represented in the data set where the underrepresented examples are added to the data set more often. This approach is chosen to feed uniformly all data for model training. This ensures that all values are selected with the same prediction probability, and the model does not act in a biased manner. 

\begin{figure}[htbp]
\centerline{\includegraphics[width=8.5cm]{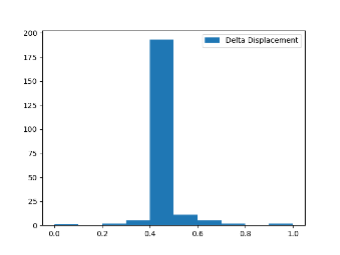}}
\caption{Non-oversampled Data Histogram (corresponding to Fig.~\ref{fig09}).}
\label{fig10}
\end{figure}

\begin{figure}[htbp]
\centerline{\includegraphics[width=8.5cm]{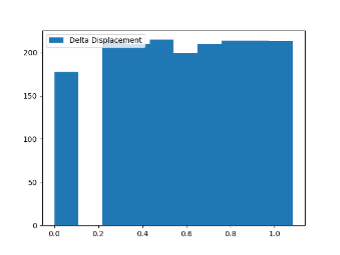}}
\caption{Oversampled Data Histogram (analog to Fig.~\ref{fig09}).}
\label{fig11}
\end{figure}

\begin{figure}[htbp]
\centerline{\includegraphics[width=8.5cm]{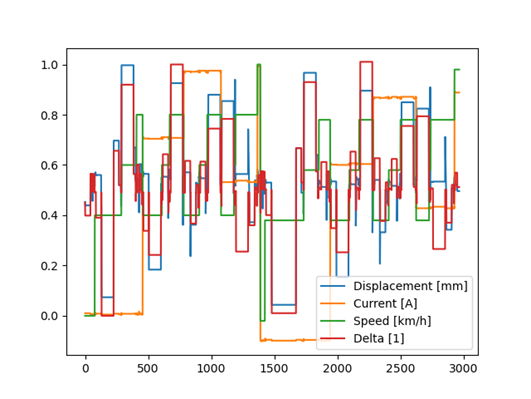}}
\caption{Completely Prepared Data Set.}
\label{fig12}
\end{figure}

Fig.~\ref{fig10} and Fig.~\ref{fig11} highlight this approach, and Fig.~\ref{fig12} is now showing the finalized train data set. Note: In Fig.~\ref{fig11} no oversampling was performed for a range of values (see red marking), since no data below a certain limit was available in the driving mode. The values close to 0 result from the stationary test.

\section{Training results}

\begin{figure}[htbp]
\centerline{\includegraphics[width=8.5cm]{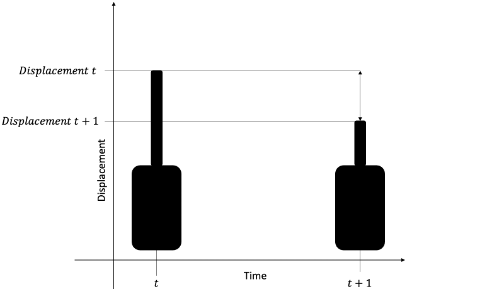}}
\caption{Explanation of Target Value.}
\label{fig13}
\end{figure}

Further the training results are discussed in relation to the prediction of the change of the piston rod position as the target value, Fig.~\ref{fig13}.

\begin{figure}[htbp]
\centerline{\includegraphics[width=8.5cm]{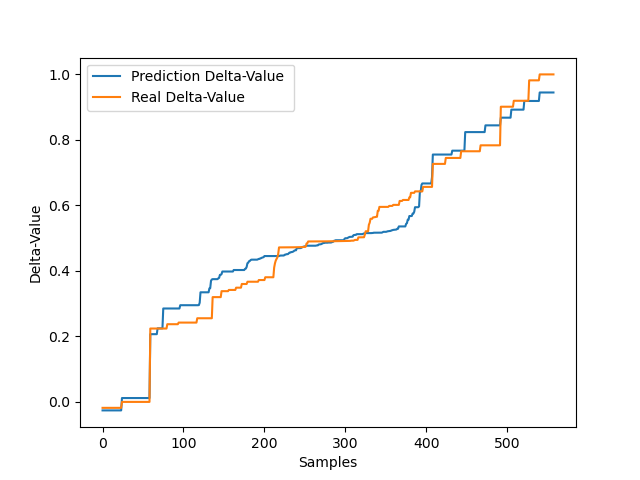}}
\caption{Test Results with Complete Preparation.}
\label{fig14}
\end{figure}

The data are now used to train the network from Fig.~\ref{fig04}. The test for previously unseen data leads to Fig.~\ref{fig14}. 

The correlating dimensionless values that quantify the model quality are shown in Table~\ref{tab2} and consist of the Mean Squared Error MSE (\ref{eq07}), the Mean Absolute Error MAE (\ref{eq08}) and the \(R^2\)-Value (\ref{eq09}). Note: Although the \(R^2\) value is generally not a suitable metric for evaluating a non-linear regression, it can provide information on the comparability of the results, therefore it was decided to include this parameter for validation purposes.

\begin{equation}
MSE=\frac{1}{N}\sum_{i=1}^{N}\left ( Y_{i}-\hat{Y_{l}}\ \right )^{2}.\label{eq07}
\end{equation}

\begin{equation}
MAE=\frac{1}{N}\sum_{i=1}^{N}\left | Y_{i}-\hat{Y_{l}}\ \right |.\label{eq08}
\end{equation}

\begin{equation}
R^{2}=1-\frac{\sum_{i=1}^{N}\left ( Y_{i}-\hat{Y_{l}}\ \right )^{2}}{\sum_{i=1}^{N}\left ( Y_{i}-\mu \ \right )^{2}}.\label{eq09}
\end{equation}

\begin{table}[htbp]
\caption{Statistical Results for Fig.~\ref{fig10}}
\begin{center}
\begin{tabular}{|c|c|c|}
\hline
\textbf{\(R^2\)} & \textbf{\(MSE\)} & \textbf{\(MAE\)}\\
\hline
0.94 & 0.01 & 0.05\\
\hline
\end{tabular}
\label{tab2}
\end{center}
\end{table}

\begin{figure}[htbp]
\centerline{\includegraphics[width=8.5cm]{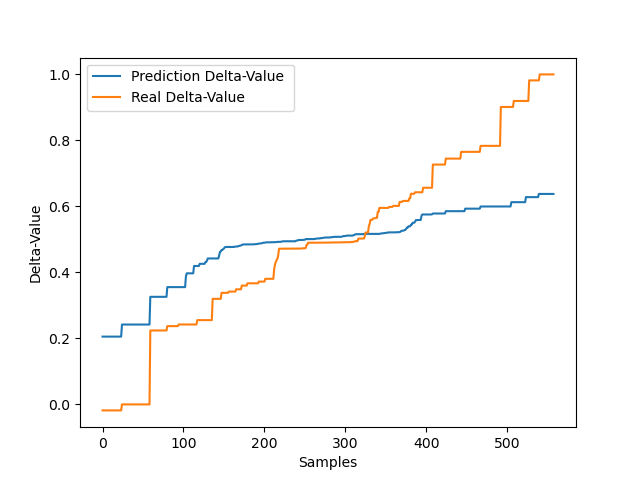}}
\caption{Test Results without Oversampling.}
\label{fig15}
\end{figure}

\begin{figure}[htbp]
\centerline{\includegraphics[width=8.5cm]{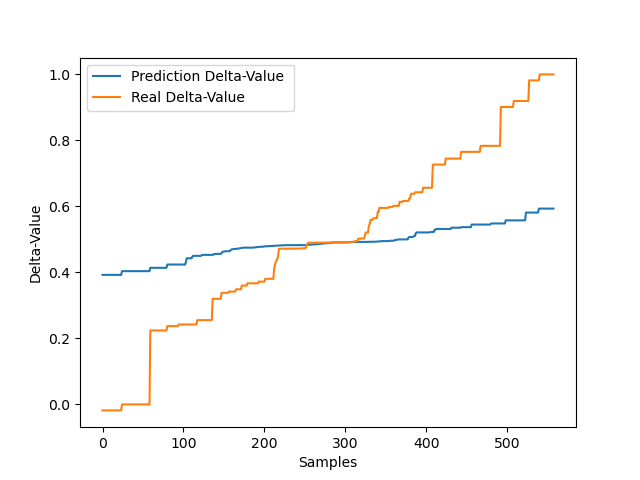}}
\caption{Test Results without Oversampling and Augmentations.}
\label{fig16}
\end{figure}

It can be seen that without oversampling the quality of the prediction decreases significantly, Fig.~\ref{fig14}, Table~\ref{tab3}. If only the augmentation method is removed, the result is being deteriorated, Fig.~\ref{fig15}, Table~\ref{tab4}. Without any pre-processing strategy, no meaningful mapping of the data course can be guaranteed Fig.~\ref{fig16}, Table~\ref{tab5}. 

\begin{table}[htbp]
\caption{Statistical Results for Fig.~\ref{fig14}}
\begin{center}
\begin{tabular}{|c|c|c|}
\hline
\textbf{\(R^2\)} & \textbf{\(MSE\)} & \textbf{\(MAE\)}\\
\hline
0.65 & 0.03 & 0.13\\
\hline
\end{tabular}
\label{tab3}
\end{center}
\end{table}

\begin{table}[htbp]
\caption{Statistical Results for Fig.~\ref{fig15}}
\begin{center}
\begin{tabular}{|c|c|c|}
\hline
\textbf{\(R^2\)} & \textbf{\(MSE\)} & \textbf{\(MAE\)}\\
\hline
0.50 & 0.04 & 0.16\\
\hline
\end{tabular}
\label{tab4}
\end{center}
\end{table}

\begin{table}[htbp]
\caption{Statistical Results for Fig.~\ref{fig13}}
\begin{center}
\begin{tabular}{|c|c|c|}
\hline
\textbf{\(R^2\)} & \textbf{\(MSE\)} & \textbf{\(MAE\)}\\
\hline
0.00 & 0.08 & 0.22\\
\hline
\end{tabular}
\label{tab5}
\end{center}
\end{table}

\begin{figure}[htbp]
\centerline{\includegraphics[width=8.5cm]{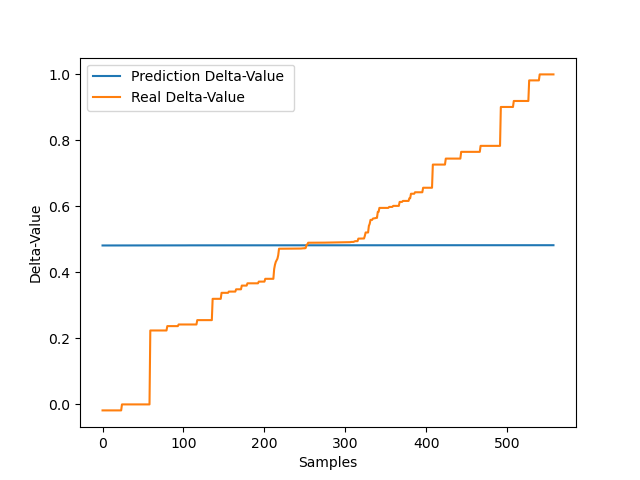}}
\caption{Test Results without Indexing, Oversampling and Augmentations.}
\label{fig17}
\end{figure}

\section{Discussion / Conclusions}
For data-driven approaches, high quality of the data is essential to ensure a valid result of the model. Especially with few data, it is highly relevant to keep the model complexity rather low (Fig.~\ref{fig04}). Otherwise, there is a risk of overfitting, which then implies advanced regularization techniques (Dropout, L2-Norm, Batch Normalization etc.). It further increases the number of hyperparameters within the model \cite{a17}, \cite{a18}.
Hence, instead of using the approach that a complex model is always needed for complex data, a sensible and well-considered pre-processing strategy is also crucial for neural networks. 

It was shown in this study that the developed approach to data preparation is sufficient even for sophisticated regression tasks. At this point it should be noted that a more complex model could not always achieve better results. Furthermore, the proposed method is suitable to generate digital twins, which represent the desired system behavior sufficiently accurate. The generated digital twin can be used in a further work to reflect the behavior of the shock absorber accurately as a training environment for the agent, where an optimal control strategy with high driving dynamics and good ride comfort can be applied by varying the current by the agent within the shock absorber. 

In addition, the model can also be used for active suspension controller parameterization, since the representation of the real system is more accurately in this way than by conventional models. A pre-simulation is also conceivable, for example, to investigate the shock absorber behavior in a protected environment in the event of an accident. Furthermore, real-time (condition-) monitoring of the shock absorber functions is also possible by comparing the behavior of the digital twin to the physical shock absorber. If this deviates from each other for some cases, this can be an indication of (i) a malfunction and therefore a damage on the shock absorber or (ii) critical wear.

\section*{Appendix}

\begin{table}[htbp]
\caption{Test program}
\begin{center}
\begin{tabular}{|c|c|c|}
\hline
Testrun & Current [A] & Velocity [km/h]\\
\hline
1 & 0.4 & 10 \\
\hline
2 & 1 & 10 \\
\hline
3 & 1.2 & 10 \\
\hline
4 & 1.5 & 10 \\
\hline
5 & 0.4 & 15 \\
\hline
6 & 1 & 15 \\
\hline
7 & 1.2 & 15 \\
\hline
8 & 1.5 & 15 \\
\hline
9 & 0.4 & 20 \\
\hline
10 & 1 & 20 \\
\hline
11 & 1.2 & 20 \\
\hline
12 & 1.5 & 20 \\
\hline
13 & 0.4 & 0 (standing still) \\
\hline
14 & 1.6 (max. current) & 25 \\
\hline
\end{tabular}
\label{tab6}
\end{center}
\end{table}

\begin{figure}[htbp]
\centerline{\includegraphics[width=8.5cm]{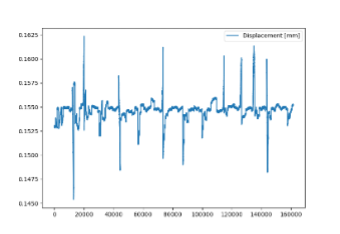}}
\caption{Original Data Unscaled, \(Displacement\).}
\label{fig18}
\end{figure}

\begin{figure}[htbp]
\centerline{\includegraphics[width=8.5cm]{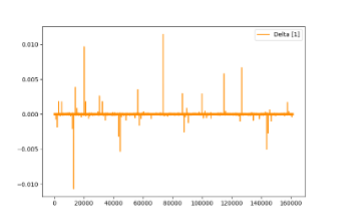}}
\caption{Original Data Unscaled, \({\Delta}_{Displacement}\).}
\label{fig19}
\end{figure}

\begin{figure}[htbp]
\centerline{\includegraphics[width=8.5cm]{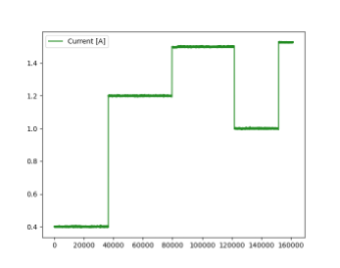}}
\caption{Original Data Unscaled, \(Current~I\).}
\label{fig20}
\end{figure}

\begin{figure}[htbp]
\centerline{\includegraphics[width=8.5cm]{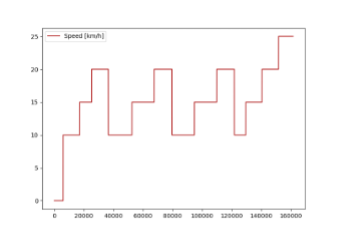}}
\caption{Original Data Unscaled, \(Vehicle~velocity~V\).}
\label{fig21}
\end{figure}

\vspace{12pt}

\end{document}